\documentclass[letterpaper,pra,aps,floatfix,superscriptaddress,onecolumn,longbibliography,notitlepage
]{quantumarticle}
\pdfoutput=1
\usepackage[utf8]{inputenc}
\usepackage[english]{babel}
\usepackage[T1]{fontenc}
\usepackage{amsmath,amsthm,amsfonts,amssymb,xcolor}
\usepackage{graphicx}
\usepackage{microtype}
\usepackage{braket}
\usepackage[numbers,sort&compress,square]{natbib}

\usepackage{subcaption}

\makeatletter
\def\l@subsubsection#1#2{}
\makeatother

\usepackage[colorlinks=true,citecolor=blue,linkcolor=blue]{hyperref}
\usepackage[capitalize,nameinlink]{cleveref}

\theoremstyle{definition}

\newcommand{\CC}{{\mathbb{C}}}

\newcommand{\diag}{{\mathrm{diag}}}

\begin{document}

\title{Boundaries for the Honeycomb Code}

\author{Jeongwan Haah}  
\affiliation{Microsoft Quantum and Microsoft Research, Redmond, WA 98052, USA}                                                                                                                                                                                                                                                              
                                                                                                
\author{Matthew B.~Hastings}                                                                                                                                                                                                                    \affiliation{Station Q, Microsoft Quantum, Santa Barbara, CA 93106-6105, USA}                                          \affiliation{Microsoft Quantum and Microsoft Research, Redmond, WA 98052, USA}                                                                                                                                                                                                                                                              

\begin{abstract}
We introduce a simple construction of boundary conditions for the honeycomb code~\cite{honeycomb} 
that uses only pairwise checks and allows parallelogram geometries at the cost of modifying the bulk measurement sequence.
We discuss small instances of the code.
\end{abstract}

\maketitle

The recently introduced honeycomb code~\cite{honeycomb} 
is a code with dynamically generated logical qubits 
that falls outside the usual stabilizer and subsystem code formalism.
This code uses only pairwise checks, 
which are a product of Pauli matrices on two different qubits.
The checks are measured in a particular sequence 
which is broken up into ``rounds'' where in each round one-third of the checks are measured, 
with the measurement pattern repeating every three rounds.

Since the honeycomb code at any moment in the dynamics is in a state 
that is virtually the same as the toric code state~\cite{Kitaev_2006,Wootton2015},
it is conceivable that a full quantum architecture 
can be built out of the honeycomb code.
One of the potential advantages of the honeycomb code over the toric code
as a basic logical element in a quantum architecture
is that the number of possible error locations per unit spacetime volume in a honeycomb code implementation
appears to be smaller than that for the toric code.
This is especially so when the codes are implemented by one- and two-qubit Pauli measurements; 
compare~\cite{Chao2020} and~\cite{Gidney2021}.
The fewer the number of ways errors might occur, the better the performance should be.

The toric code has well-known boundary conditions~\cite{Bravyi1998,Freedman1998} (rough and smooth~\cite{Dennis})
with which one can implement a version of the code, called the surface code,
using nearest-neighbor interactions on a two-dimensional planar grid of qubits.
This is quite important
because of a straightforward layout of many logical qubits in a plane,
making the surface code appealing for a quantum architecture at scale.
In contrast, the honeycomb code's boundary conditions were relatively poorly understood.
The dynamics interchanges electric and magnetic operators after every round,
and hence, in order to construct a code with boundaries rather than on a torus,
it is necessary for the rough and smooth boundary conditions to alternate every round.
To this end, we sketched~\cite{honeycomb} one solution
where the code is shrunk after certain measurement rounds to preserve the correct boundary conditions, 
and then periodically grown by using some nonpairwise checks.
However, a simpler solution is certainly desirable.
In \cref{sec:bc}, 
we present a simple solution for boundary conditions of the honeycomb code
which involves modifying the bulk measurement sequence.
In \cref{sec:surgery}, we discuss surgery.
In \cref{sec:smallcode}, we briefly consider tilings of the torus to construct small codes and consider their distance.
A detailed performance comparison will require numerical simulation~\cite{sim}.

While this paper was in preparation, 
another planar realization of the honeycomb code was given~\cite{Vuillot2021}.
Unfortunately, the boundary conditions of~\cite{Vuillot2021}
make the protocol non-fault-tolerant.
We will give some intuition for this at the end of~\cref{sec:decodingGraphs}.

\section{Boundary conditions}
\label{sec:bc}
We begin by considering ways of ``gapping'' the instantaneous stabilizer group (ISG) with boundaries in \cref{2gon4gon}.
Then in \cref{modbulk}, we describe a modified bulk measurement sequence to realize a dynamically generated logical qubit 
on an annulus (or more generally on a multiply punctured disk).
Finally, in \cref{parallel}, we give a parallelogram geometry.

\subsection{2-gon and 4-gon boundaries}
\label{2gon4gon}

\begin{figure}
\centering
\includegraphics[width=\textwidth, trim={0ex 70ex 0ex 0ex}, clip]{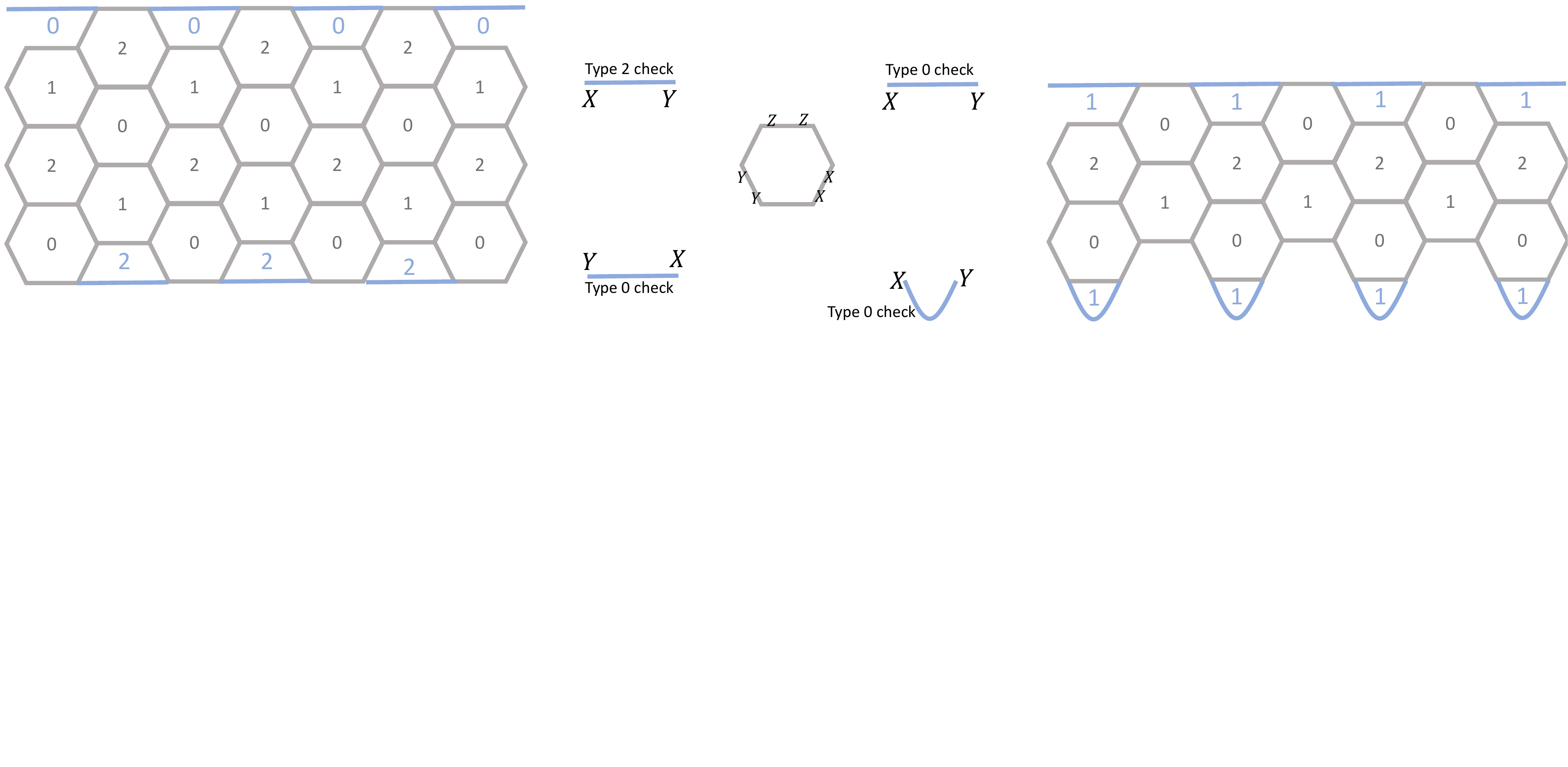}
\caption{
	Left: armchair with 4-gons on both boundaries.  
	Right: armchair with 2-gons on bottom boundary and 4-gons at top.
	These two figures show that we can tune the height of the strip by using different boundary conditions.
	We have displayed the boundary check operators explicitly.
}
\label{fig:2gon4gon}
\end{figure}

The left side of \cref{fig:2gon4gon} is the same as Fig.~7 of~\cite{honeycomb},
with the addition of some extra cells at the top to gap both boundaries.
This shows one way to introduce boundaries. 
Assume the left and right edges are continued further and then joined to make an (topological) annulus.
The geometry of the hexagons may be termed an ``armchair'' geometry at the boundary,
following terminology in carbon nanotubes. 
We have added additional edges to create additional plaquettes with four edges, that we term \emph{4-gons}.
Each 4-gon is a type~2 plaquette at the bottom, and the types of the edges are defined correspondingly:
an edge at the very bottom that belongs to a hexagon is type~2,
and an edge at the very bottom that belongs to a 4-gon is type~0.
The void below the lattice can be regarded as a very large plaquette of type~1.
At the top, type~0 and type~2 are interchanged; 
the void above the lattice is regarded type~1.

The right side of \cref{fig:2gon4gon} shows an alternative.
In this case we have added \emph{2-gons} (also sometimes called bigons) at the bottom, 
each of which has only two edges.
These 2-gons are of type~1 in this case.
Each added half-circle edge is type~0,
and the other edge of a 2-gon is type~2.
Note that the perpetual plaquette stabilizer on a 2-gon has weight~$2$.
Further, notice that if we measure a check which is an edge of a 2-gon (either type~0 or type~2 in this figure),
then the two qubits in the 2-gon are in a Bell state, and so disentangled from the others.

\subsection{Embedded surface codes}

It is routine~\cite{honeycomb} but is crucial to our construction
to identify an embedded surface code state with gapped boundary after each round.
Every check projects a pair of qubits into an effective qubit ($\CC^2$),
and one can draw a line segment orthogonal to the edge that covers the pair.
These line segments give a superlattice of effective qubits,
and perpetual plaquette stabilizers correspond to 
supervertex and superplaquette stabilizers on the effective qubits.
The result is drawn in~\cref{fig:stripDynamics44,fig:stripDynamics42};
see the quadrants labeled by 0, 1, and 2;
an exception is given to step~$0^*$ 
where an effective qubit may be due to a surviving check operator from the previous round.

In fact, there is a canonical choice of Clifford frame for each effective qubit,
implied by the superlattice stabilizers.
Here, a Clifford frame is nothing but a choice of logical operators (up to a sign)
on the $[[2,1,1]]$ code defined by the check operator.
Each superplaquette contains exactly one plaquette of the original lattice,
on which the perpetual stabilizer acts by some single-qubit Pauli operator on each effective qubit
that surrounds the superplaquette. 
That single-qubit operator must be identified with $Z_\text{eff}$ on the effective qubit.
Likewise, each supervertex is contained in exactly one original plaquette,
on which the perpetual stabilizer is a tensor product of several two-qubit operator on edges,
which we identify with $X_\text{eff}$ up to a sign.
For $Z_\text{eff}$ in the bulk, a simpler rule is that 
if a check operator is $a \otimes b$, 
then $Z_\text{eff} = \pm (a \otimes I) = (I \otimes b)( a \otimes b)$.

The thickness of the annulus is chosen 
such that the top boundary conditions of the embedded toric code state match the bottom ones.
With a different thickness it is possible to have 2-gons at both top and bottom.

\subsection{Modification of bulk sequence}
\label{modbulk}

\begin{figure}
\centering
\includegraphics[width=\textwidth, trim={0ex 0ex 0ex 7ex}, clip]{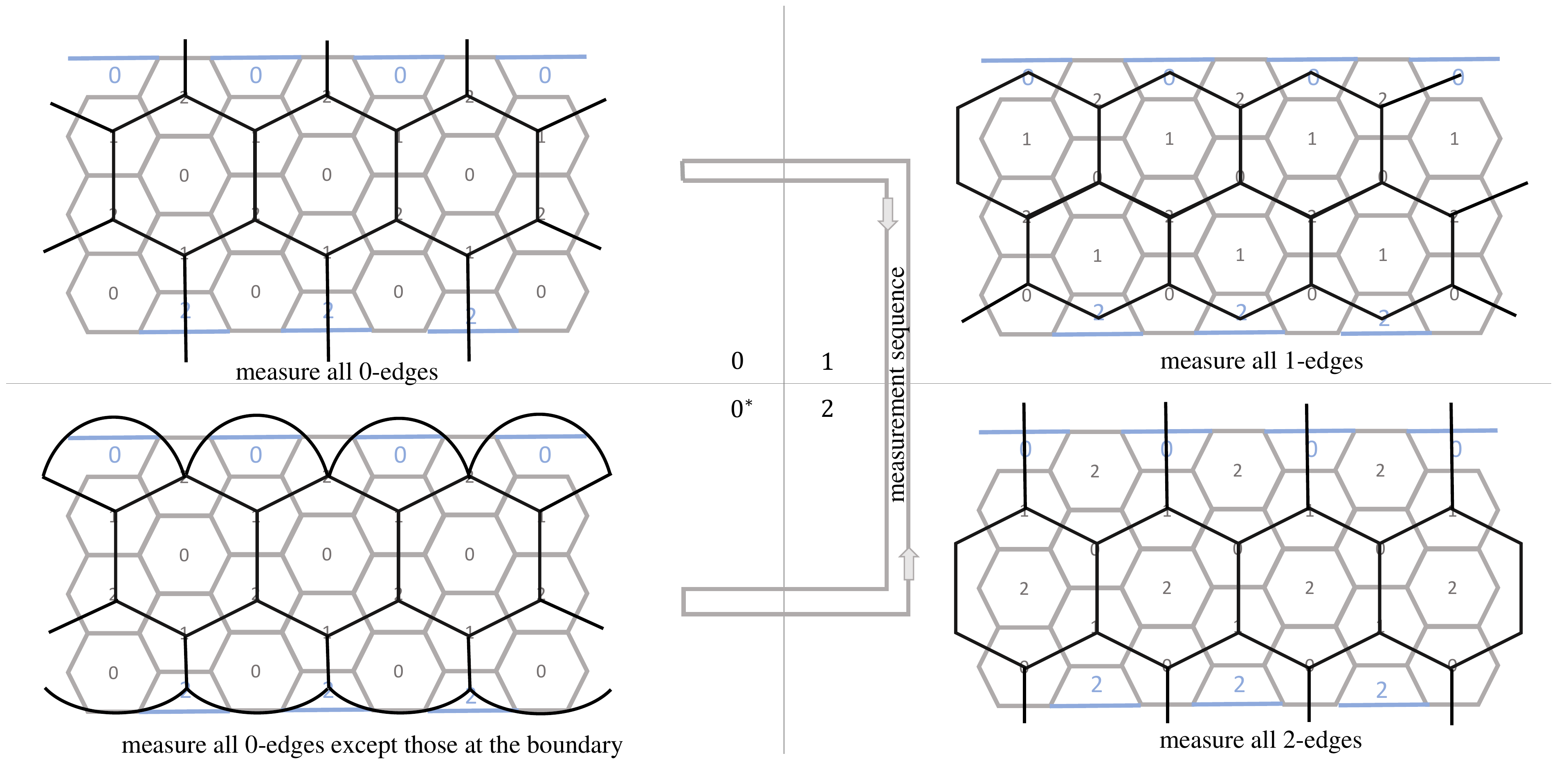}
\caption{
	At each round, the superlattice supports an embedded toric code state with boundaries.
	\emph{After round~0},
	each 4-gon at the bottom has two edges of type~0,
	both of which are measured.
	Since there is a perpetual plaquette stabilizer at each 4-gon,
	the instantaneous stabilizers on a 4-gon 
	project its four qubits down to $\CC^2$,
	depicted by a slightly longer vertical superedge.
	\emph{After round~2}, 
	a similar projection is realized at the top 4-gons of type~0.
	\emph{After round~0$^*$},
	the edges of type~0 at the boundary (top and bottom) are \emph{not} measured, 
	and therefore the type~2 check at the boundary remains in the ISG, 
	projecting its two qubits to $\CC^2$.
	These projections make super-3-gons and super-5-gons at the boundaries.
}
\label{fig:stripDynamics44}
\end{figure}
\begin{figure}
\centering
\includegraphics[width=\textwidth, trim={0ex 0ex 0ex 10ex}, clip]{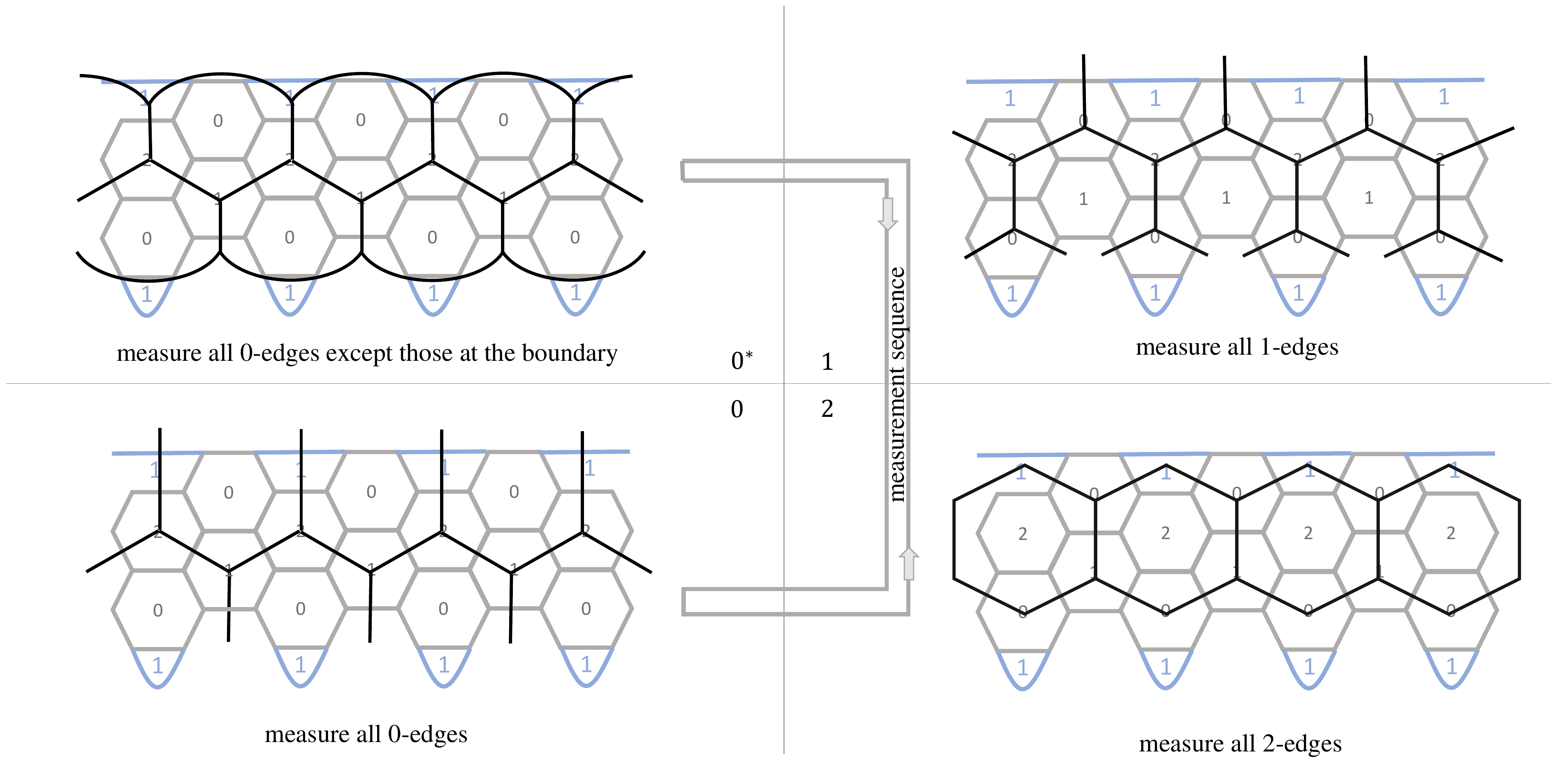}
\caption{
	Superlattices similar to those in \cref{fig:stripDynamics44}.
	\emph{After round~0$^*$},
	the checks of type~1 at the top boundary remain in the ISG,
	rendering the superlattice to have 5-gons at the top boundary.
	The perpetual plaquette stabilizer on each 2-gon
	projects the two qubits of the 2-gon to $\CC^2$,
	rendering the superlattice to have 5-gons at the bottom boundary.
	The product of the two type~1 checks positioned the lowest in the figure around a ``bay,''
	survives in the ISG and gives the supervertex stabilizer.
	\emph{After round~2},
	the type~2 check on the edge of a 2-gon and its perpetual stabilizer
	make a Bell state, disentangling the two qubits of the 2-gon from the others.
	\emph{After round~0},
	each 4-gon makes its four qubits into one effective qubit,
	and each 2-gon makes a Bell state.
}
\label{fig:stripDynamics42}
\end{figure}

Under the choice of boundaries with all 4-gons as in the left side of \cref{fig:2gon4gon},
if we measured type~2 checks and then type~0 checks,
we would reveal the inner logical operator, destroying the dynamically generated logical qubit.
This may be understood in two ways.
One way is that the bulk dynamics interchanges electric and magnetic operators every round, 
while the boundary conditions after round~2 and round~0 
are of the same type (both rough) rather than alternating.
Another way is to view the annulus on a topological $2$-dimensional sphere
so that the bottom edges form the boundary of a single ``very large plaquette'' of type~1;
measuring type~2 and then type~0 checks infers the stabilizer of the very large plaquette.
A similar problem arises with the right side of \cref{fig:2gon4gon}.
In this case, the ``very large plaquette'' is of type~2,
whose stabilizer (an inner logical operator) would be revealed
if we measured type~0 and then type~1 checks.

To have a dynamically generated logical qubit on an annulus, 
while using only pairwise checks, 
we modify the bulk measurement sequence as follows.
We measure edges of type
\begin{equation}
	0,1,2,0^*,2,1
\end{equation}
in order and repeating.
Depending on the boundary geometry we may use a sequence
\begin{equation}
	0^*,1,2,0,2,1.
\end{equation}
They have period~$6$, not~$3$.
The notation is that $0,1,2$ means measuring all checks of the given type,
but $0^*$ means measuring all type-0 checks
{\it except} those on a boundary.
Thus we do not reveal the inner operator.

Both sequences infer all small plaquettes:
for $(a,b,c)$ that is a permutation of $(0,1,2)$,
after measuring checks of type~$a$ and type~$b$ in succession,
we infer plaquette stabilizers of type~$c$.
In other words, we infer plaquette stabilizers of type $2,2,0,1,1,0$ in sequence.
This is independent of where we have $0^*$ in the measurement sequence.
See \cref{fig:stripDynamics44,fig:stripDynamics42}.
The boundary conditions in the embedded toric codes are alternating from round to round as they should.


\subsection{Parallelogram}
\label{parallel}

\begin{figure}[h]
\centering
\includegraphics[width=\textwidth]{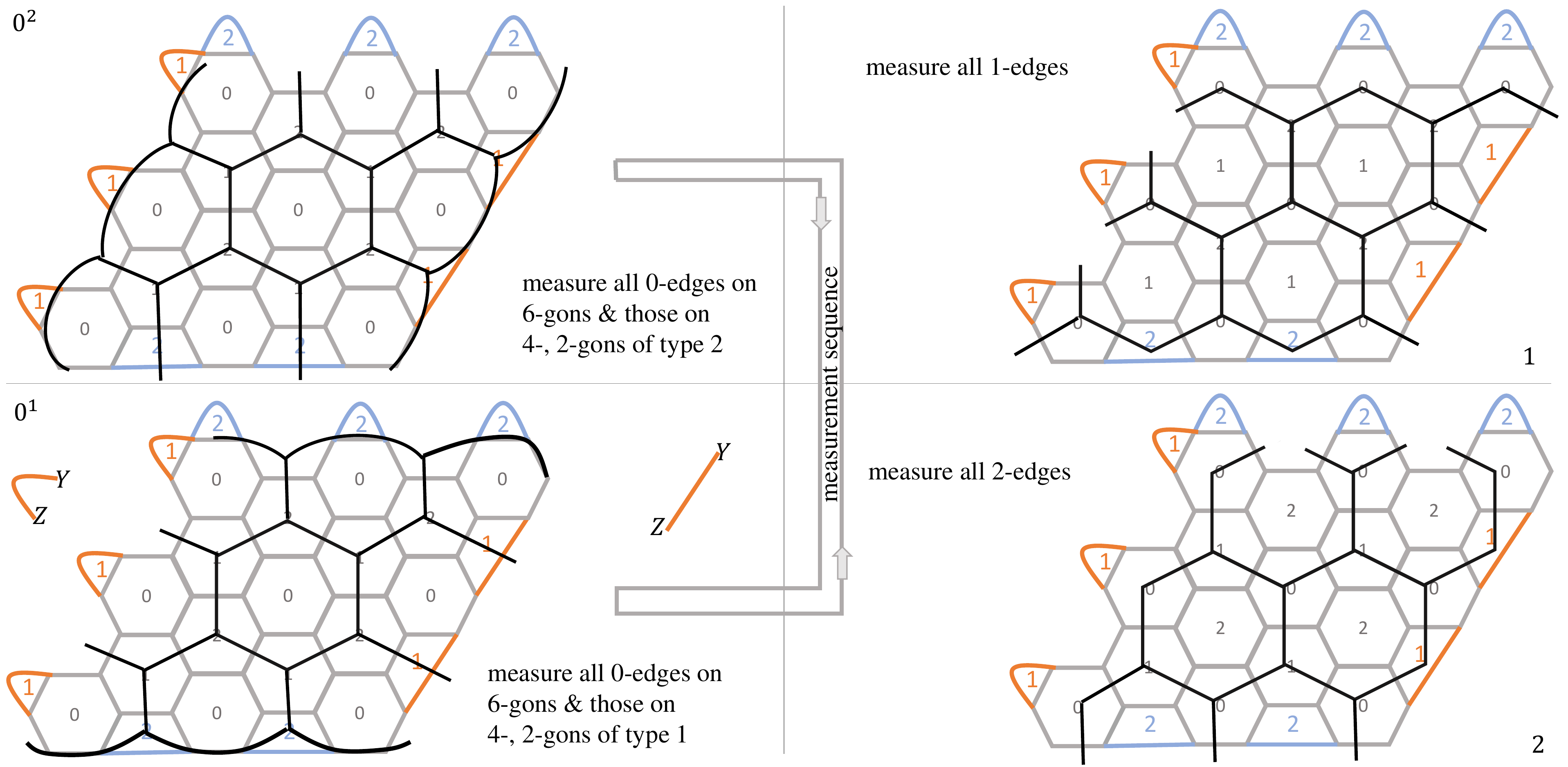}
\caption{
A parallelogram patch of honeycomb code using 4- and 2-gon boundaries.
The construction of superlattice follows the same reasoning as in \cref{fig:stripDynamics44,fig:stripDynamics42}.
New phenomena occur at the corners.
First, the top left qubit participates in four checks,
and there is no static plaquette stabilizer associated with the two intersecting 2-gons;
however, the three qubits in the support of these two 2-gons
are in some subspace $\CC^2$ always (except for the very first round).
Second, in the lower left corner after round~0$^2$
the perpetual plaquette stabilizer at the 2-gon
and the type~1 check from the previous round~1 
project the three qubits covered by the curvy line to a subspace $\CC^2$.
A symmetric situation (1$\leftrightarrow$2) is observed in the top right corner at round~0$^1$.
}
\label{fig:p24}
\end{figure}

To encode a single logical qubit on a parallelogram in the usual toric code, 
one pair of opposite sides (top \& bottom) of the parallelogram 
should have smooth boundary conditions, 
while the other pair (left \& right) should have rough boundary conditions.
In the honeycomb code, 
we have to make sure that boundaries of a parallelogram alternate in time
while neighboring sides have opposite boundary conditions.
One solution is depicted in \cref{fig:p24}.
It is straightforward to make the parallelogram larger.
Note that the patch depicted in \cref{fig:p24}
can fill the infinite plane of the honeycomb lattice of qubits
with no gaps in between parallelograms.

A strategy to construct a parallelogram with armchair boundaries is as follows.
(i)~We choose the height of a horizontal strip and use 2- or 4-gon boundaries (as in \cref{fig:2gon4gon})
such that the top and bottom boundaries have the same type of superlattice boundaries (rough and smooth)
at round~1 and~2.
(ii)~We truncate the strip on the left and use 2- or 4-gon boundary geometry
such that the superlattice boundary condition on the left
is opposite to that on the top and bottom at round~1 and~2.
(iii)~We truncate the strip on the right such that the superlattice boundary condition 
is the same as that on the left.
Whether we use 2-gon or 4-gon boundary geometry on the right 
will be dictated by the width of the parallelogram.
(iv)~By demanding the spatial alternation of superlattice boundary conditions along the boundary,
we determine which kind of round~0 should be used in the measurement sequence
\begin{equation}
	0',1,2,0'',2,1
\end{equation}
where 0$'$ or 0$''$ indicates that certain type~0 checks are not measured at the boundaries at that round.

\subsection{More general shapes}

A general prescription to build a planar patch that encodes one logical qubit is as follows.
A qubit must participate in exactly $3$ checks except for four qubits
where the boundary condition of the embedded surface code changes.
In each portion of the boundary on which the qubits have degree~$3$ only,
the types of the edges should alternate between two options.
The bulk measurement sequence is still $\ldots, 0,1,2,0,2,1,\ldots$ that has period~$6$.
At round~$1$ and~$2$, all checks of a given type is measured.
At round~$0$, some type~$0$ checks near the boundary are omitted,
which can be easily identified by demanding 
that the embedded surface code should have time-alternating boundary conditions.
In~\cref{fig:p24} there are three qubits that are acted on by only two checks,
and one qubit that is acted on by four checks.
In~\cref{fig:logZX} below we construct a patch with two degree-$3$ qubits,
and two degree-$4$ qubits.

\subsection{Decoding graphs}\label{sec:decodingGraphs}

\paragraph{Bulk.}
We begin by considering the infinite honeycomb lattice with the period~6 sequence, $0,1,2,0,2,1$.
Since there is no boundary, the round~0 is not modified.
We continue to use the simplified model~\cite{honeycomb}
in which each check $E \otimes E$ of weight~$2$ at any location in spacetime 
may be followed by an error $E$ on one of the qubits,
but no other error occurs.
This simplified error model generates 
all possible error patterns including measurement outcome flips.
Type~$a$ checks at a round and type~$b$ checks at the next round
infer type~$c$ plaquette stabilizers, where $a,b,c$ are distinct.
So, the sequence of plaquette types that are inferred (immediately after the check operators are measured) is
\begin{equation}
2, 0, 1, 1, 0, 2
\end{equation}
that has period~$6$.

An error $E$ after a type~$a$ check and before a type~$b$ check,
flips a fresh type~$c$ plaquette stabilizer;
by construction, $E$ anticommutes with the type~$b$ check that follows $E$ in time.
All subsequent plaquette stabilizers are inferred by checks that are measured after $E$,
and therefore whether a given plaquette stabilizer $P$ is flipped by $E$ 
is determined by the commutation relation between $E$ and $P$.%
\footnote{
	This conclusion is absurdly trivial in a usual stabilizer code.
	However, in our consideration, 
	some errors (that are not first order in the current simplified error model)
	that commute with all plaquette stabilizers,
	can cause nontrivial syndromes 
	due to the timing when the plaquette stabilizers are inferred.
	In other words, some error that commutes with a plaquette stabilizer
	may still excite the syndrome node associated with the plaquette stabilizer
	if the error is sandwiched between the two measurements in time,
	by which the eigenvalue of the plaquette stabilizer is inferred.
}
There are exactly two plaquette stabilizers that anticommute with $E$.

We examine the plaquette stabilizers in spacetime that are inferred \emph{after} a given error
and mark them with $\star$ if they are the \emph{first} flip of the plaquette stabilizer.
Due to the spatial symmetry, we only have to look at the plaquette types.
\begin{align}
\begin{array}{c|llllll|llllll|c}
E \text{ after round } 0 &2\star&0&1\star&1&0&2 &2&0&1&1&0&2 &\cdots\\
E \text{ after round } 1 & &0\star&1&1&0&2\star &2&0&1&1&0&2 &\cdots\\
E \text{ after round } 2 & & &1\star&1&0\star&2 &2&0&1&1&0&2 &\cdots\\
E \text{ after round } 0 & & & &1\star&0&2\star &2&0&1&1&0&2 &\cdots\\
E \text{ after round } 2 & & & & &0\star&2 &2&0&1\star&1&0&2 &\cdots\\
E \text{ after round } 1 & & & & & &2\star &2&0\star&1&1&0&2 &\cdots
\end{array}
\end{align}
where $E$ is a single-qubit error that immediately follows a check at the indicated round.
This table allows us to construct a decoding graph.
The nodes are the plaquette stabilizers in spacetime, 
whose time coordinates are when they are inferred.
A node indicates if the associated plaquette stabilizer has an eigenvalue
that is \emph{different} from that of the same plaquette immediately preceding in time.
An error (in the current simplified noise model) causes flips in the nodes indicated by $\star$,
which happen to be two in number,
so the error gives an edge between those flipped nodes.

We see that there is no edge between time slices of an odd time coordinate difference.
If we collect all time slices with a fixed time coordinate parity,
the decoding graph is connected.
Hence, there are two separate decoding graphs.
They are isomorphic due to the symmetry 1$\leftrightarrow$2.
The decoding subgraph is not regular.
Recalling the $2\pi/3$ rotation symmetry of the honeycomb lattice,
we see that in the subgraph containing nodes of the first column in the table,
type~2 nodes have degree~$3$,
type~0 nodes have degree~$6$,
and
type~1 nodes have degree~$9$ ($6$ to the past and $3$ to the future).
The same matching graph argument as in \cite{honeycomb} proves that 
there is a fault tolerance accuracy threshold for this model that is nonzero.
Uncorrectable errors are precisely those that give a nontrivial homology cycle in the decoding graph.


\paragraph{Annulus.}

With boundaries at the top and bottom of a strip,
there is a new type of syndrome bit.
From the decoding graph of the infinite bulk,
the decoding graph for a topological annulus
can be obtained with the following modifications.
\begin{itemize}
	\item (Perpetual stabilizer) Keep bulk nodes that correspond to existing plaquettes and edges among them.
	\item (Transient stabilizer)
		Some check~$c$ of type~$a$ ($a=1,2$) at the boundary near a $4$-gon 
		survives at round~$0$ that is sandwiched by two rounds of type~$a$
		because we do not measure certain type-$0$ edges.
		Similarly, near a $2$-gon, the product~$c'$ of the two type-$a$ checks 
		on the ``armrest'' of the ``armchair''
		survives at round~$0$.
		Such a check~$c$ or a product~$c'$ defines a transient stabilizer 
		whose value is remeasured at the subsequent round~$a$.
		The consistency of these values is a syndrome bit.
		It is important to keep track of these transient stabilizers,
		since they contribute to the boundary conditions of the decoding graph,
		giving correct homology for one logical qubit.
	\item Add a terminal node for each fictitious ``very large plaquette'' beyond the boundaries.
	\item Connect the terminal nodes to the bulk nodes,
	as if it were a usual plaquette neighboring the boundary plaquettes.
	The terminal node has high degree.
\end{itemize}
Uncorrectable errors are those that give a nontrivial homology cycle \emph{relative} to the terminal nodes.
The connected component of the decoding graph which contains the new type of syndrome bits,
has smooth boundaries and is disconnected from the terminal nodes.
The other subgraph has rough boundaries if we delete the terminal nodes.
This is analogous to the surface code on a topological annulus
where one component of the decoding graph accounts for $X$ errors,
and the other for $Z$ errors.

\paragraph{Non-fault-tolerance due to boundaries.}

In introduction, we have remarked that the boundary construction 
of~\cite{Vuillot2021} renders the code non-fault-tolerant.
This phenomenon may seem subtle because every ISG in the scheme of~\cite{Vuillot2021}
defines a code with a large code distance.
However, a constant-weight fault configuration leading 
to a logical error exists in spacetime, 
corresponding to the motion of an anyon with \emph{finite} velocity.
In~\cite{Vuillot2021},
the period~$3$ measurement sequence is used for a whole planar patch.
In particular, the ISG with boundary included has period~$3$ up to signs.
Consider an $e$-particle~$x$ that enters into the bulk from a boundary component 
at time~$t$.
Suppose that $x$ stays in the bulk for $3$ measurement rounds.
Due to the nontrivial automorphism implemented by the measurement sequence,
the particle~$x$ becomes~$m$ at time~$t+3$.
However, the boundary conditions of the ISG at time~$t+3$,
through which $x$ has entered, is the same as those at time~$t$,
so the particle may exit, not through the same spatial location,
but only through a different boundary component.
The whole process may occur at a corner 
where the boundary conditions of ISG changes,
and therefore can be implemented by a constant-weight operator in spacetime.
On the other hand, nothing prevents for an $m$-particle~$y$ at time~$t$
to stay near the corner indefinitely. 
The worldline of~$y$ can therefore link with that of~$x$,
suggesting that $x$ may implement a logical error.
Here, it is necessary that $x$ stays undetected to any decoder.
It turns out that this is indeed possible,
as not all stabilizers are inferred every measurement round.

\paragraph{Parallelogram.}

The prescription for the annulus case applies verbatim;
there are transient stabilizers along the boundary,
which are inferred whenever the check measurement outcome 
is deterministic in the absence of errors.
Note that in contrast to~\cite{Vuillot2021},
our boundary conditions evolve in time in a commensurate way
with the nontrivial anyon automorphism in the bulk.
Any anyon that enters through a boundary component~$b$ may exit
through either $b$ or the opposite boundary component~$b$.

\section{Surgery}\label{sec:surgery}

In this section we consider logical operations across neighboring parallelograms.
We present protocols to 
embed a qubit into a larger patch,
and to measure a product of logical operators, 
$\bar X \bar X$, $\bar Z \bar Z$, $\bar X \bar Z$, or $\bar Z \bar X$, 
across two patches.
Since we know how to prepare an eigenstate of logical $\bar X$ or $\bar Z$
and to measure a logical patch in~$\bar X$ and~$\bar Z$ basis~\cite{honeycomb},
if supplemented with a single-qubit logical gate $S = \diag(1,i)$ or state~$\ket 0 + i \ket 1$,
our measurement portfolio implements all Clifford operations.

Indeed, the joint $\bar X\bar X$ or $\bar Z \bar Z$ measurement, 
the single-qubit preparation in $\bar X$ or $\bar Z$ basis,
and the single-qubit measurement in $\bar X$ or $\bar Z$ allow quantum teleportation.
Logical CNOT is possible by the joint and single-qubit measurements~\cite[\S5]{Gottesman1998}.
The group of single-qubit Clifford operations is generated by~$S$ and the Hadamard gate~$H$.
The $S$ gate is implemented by injecting $\ket 0 + i \ket 1$ 
for which we need the joint $\bar Z \bar Z$ and single-qubit $\bar X$ measurements.
The effect of~$H$ is to interchange~$\bar X$ and~$\bar Z$,
but our measurement portfolio is closed under such an interchange.
Therefore, we achieve Clifford completeness.

\subsection{Growing patches}

This step is used to prepare a patch containing a magic state~$\ket 0 + e^{i\pi/4} \ket 1$.
The prescription for a usual surface code~\cite{Li2014} 
is that we prepare data qubits that are going to be occupied by a larger patch
in an eigenstate of the to-be-grown part of the logical operators,
and begin stabilizer measurements.
Since the logical operators are conserved operators of the stabilizer measurements,
the grown patch will have the original logical information encoded.
The same strategy is applicable for our honeycomb code.
We prepare ancillas in an eigenstate of the to-be-grown part of the instantaneous logical operator,
and begin our check measurement dynamics.

\subsection{Logical $XX$ and $ZZ$ measurements}

We bring two regular patches side by side.
Then, we stop the measurements for the patches and start the measurement sequence 
for the double sized patch.
The boundary 2-gons and 4-gons are dropped in between the patches
and the ``gap'' between the patches is filled with bulk hexagons and one 4-gon.
The logical outcome of the surgery is given by the product of the new plaquette stabilizers.
Whether the surgery implements logical $\bar X \otimes \bar X$ or $\bar Z \otimes \bar Z$ measurement
depends on whether the relative displacement of the two patches is horizontal or vertical.

\subsection{Logical $XZ$ and $ZX$ measurements}

\begin{figure}
\centering
\includegraphics[width=\textwidth]{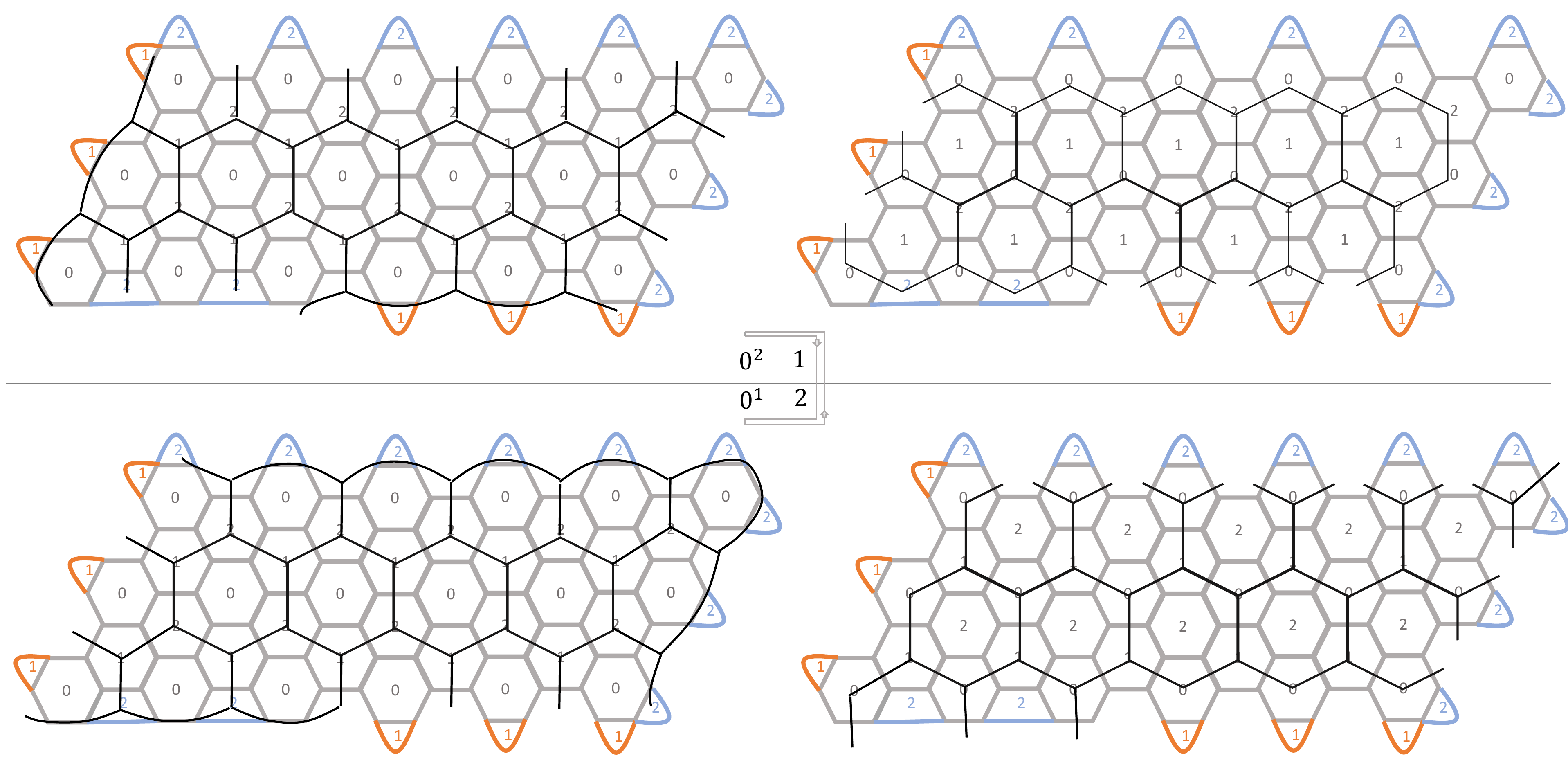}
\caption{
	An extended patch.
	The displayed patch occupies space that is worth two parallelograms.
	Observe that along the bottom boundary
	there are both smooth and rough sections.
	At the top right corner the type~0 hexagon 
	is disentangled from the rest in round~1 and~2.
}
\label{fig:extpatch}
\end{figure}

\begin{figure}[p]
\centering
\includegraphics[width=\textwidth, trim={0mm 10mm 0mm 0mm}, clip]{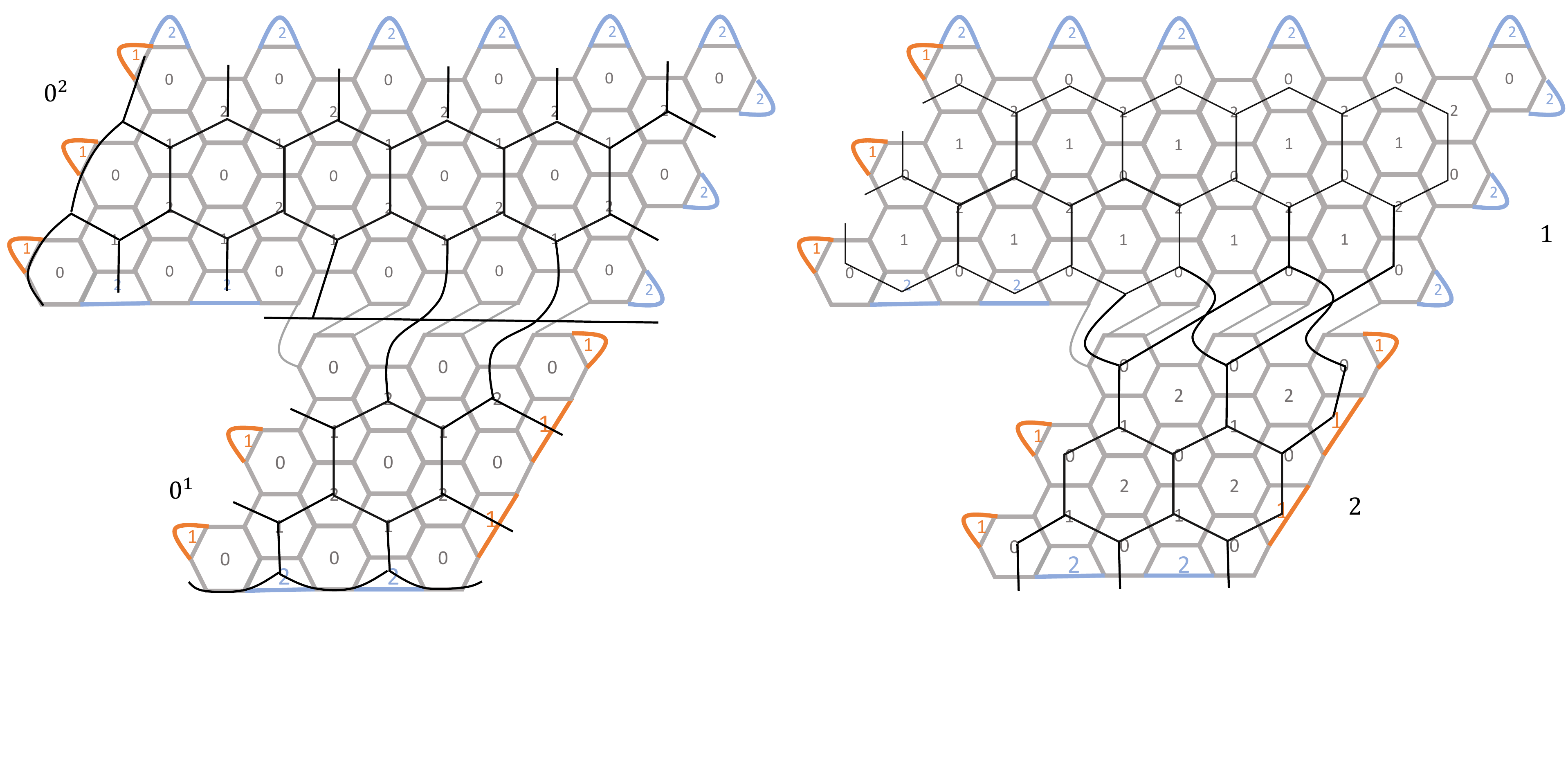}
\includegraphics[width=\textwidth, trim={0mm 10mm 0mm 0mm}, clip]{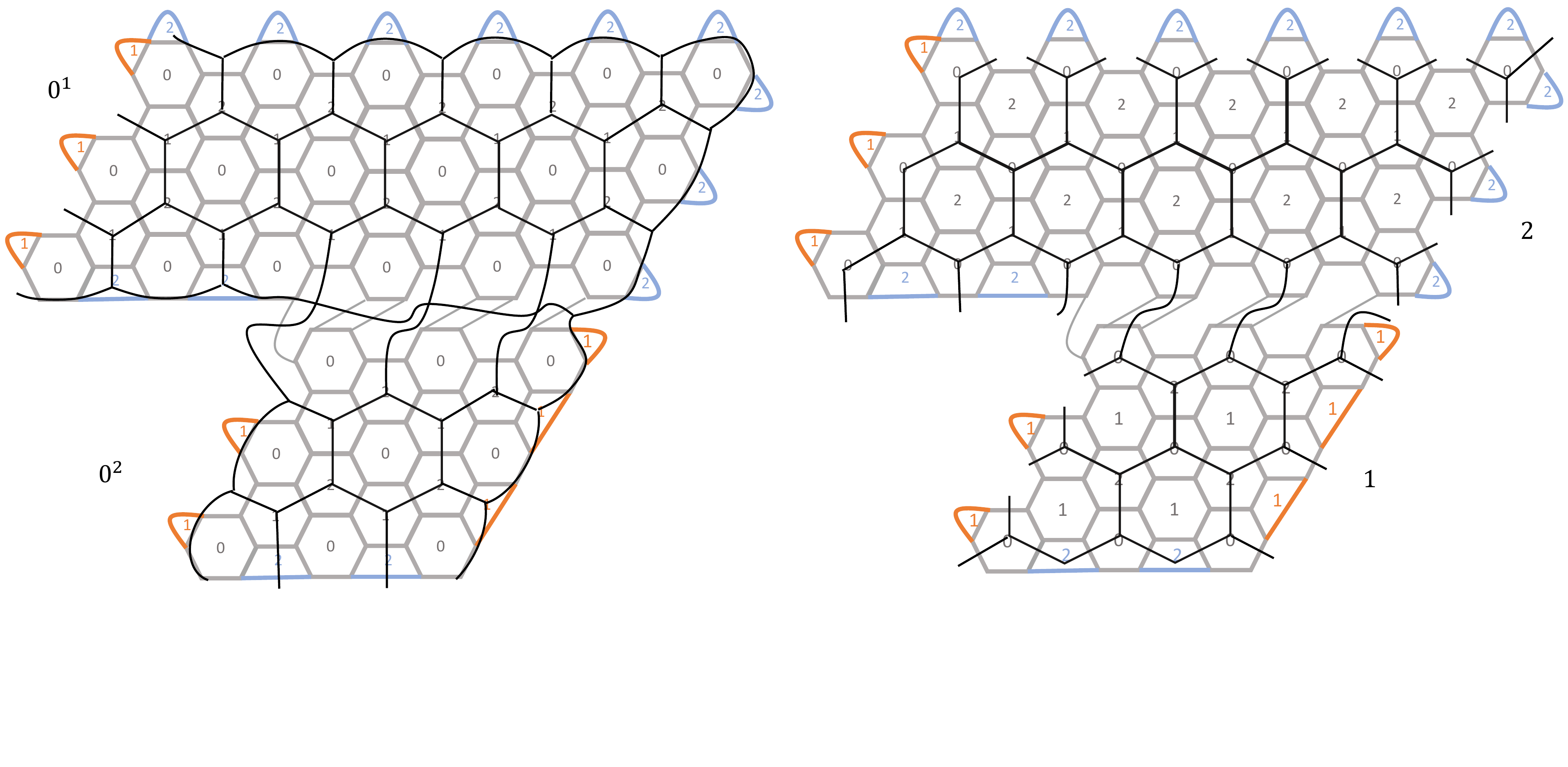}
\caption{
	An extended patch of \cref{fig:extpatch} is stitched with a parallelogram patch.
	The qubits form a portion of honeycomb lattice and 
	the hexagon labels are inherited from the infinite honeycomb lattice.
	Unlike other surgery operations, the upper extended patch and the lower regular patch
	run measurement schedules that are opposite of each other,
	which is equivalent to running the same sequence but with offset of 3 rounds.
	This offset synchronization can be implemented
	by letting one of the patches	idle for $3$ rounds.
	At the stitch, all the new edges are type~0, each of which is measured at round~0.
	The outcome of the logical measurement is given by the product of 
	the plaquette stabilizers that depend on new type~0 edges.
	If this surgery measures logical $\bar Z \otimes \bar X$,
	the symmetric version where the extended patch is extended vertically
	measures logical $\bar X \otimes \bar Z$.
}
\label{fig:logZX}
\end{figure}

We bring two regular patches 
so that they are diagonally separated in the square lattice of regular patches.
Then we merge the two patches via lattice surgery depicted in \cref{fig:extpatch,fig:logZX}.
While the extension in \cref{fig:extpatch} and the merger in \cref{fig:logZX} do not have to be done separately,
it is conceptually easier to understand the full surgery by regarding it in those two steps,
so we have drawn two sets of figures.
The extension step is no different from the usual extension above.
The merger is also similar;
we stop the measurement sequences for each patch
and begin the depicted measurements.
The measurement sequences after the merger for the upper part and the lower part
are opposite of each other:
\begin{align}
\begin{array}{cccccccc}
\text{upper:} &0^2&1&2&0^1&2&1& \cdots \\
\text{lower:} &0^1&2&1&0^2&1&2& \cdots
\end{array}
\end{align}
This opposite sequence is introduced 
to cope with the time-alternating boundary conditions of the embedded surface code;
e.g., at round~$1$, the embedded surface code on the parallelogram 
has the smooth boundary at the top,
but the rough boundary at the lower bottom of the extended patch.
It remains open if the merger is possible using two-qubit Pauli measurements only without introducing the opposite measurement sequence.
The opposite sequence can be most easily implemented 
by letting one of the patches idle for 3 rounds.
For example,
\begin{align}
\begin{array}{ccccccccccccccc}
\text{upper:}&\cdots &0^2&1&2&0^1&2&1&0^2&1&2&0^1&2&1& \cdots \\
\text{lower:}&\cdots &0^2&1&2&   & & &0^1&2&1&0^2&1&2& \cdots
\end{array}
\end{align}
This is likely the best strategy if the idling qubit noise is low compared to operation noise.
Since there is no type~1 or~2 edge crossing the stitch,
the type of each edge is unambiguously determined by that of the pre-merger patches.

\section{Small codes on torus}
\label{sec:smallcode}

In general, the honeycomb code can be defined for any trivalent graph $G$ for which the faces can be $3$-colored.%
\footnote{
	Of course, if the graph is not a honeycomb, then the name ``honeycomb code'' may be less appropriate.
}
This is the same condition needed to construct a color code~\cite{bombin2006topological}.
Examples include the $4.8.8$ and $4.6.12$ uniform tilings.

Since $G$ is trivalent, every face of the dual graph $G^*$ is a simplex (i.e., a triangle).
The $3$-coloring of the faces of $G$ gives a $3$-coloring of vertices of $G^*$.
Excluding some degenerate examples in which $G^*$ does not define a simplicial complex,
the dual $G^*$ is a triangulation of the torus.
The smallest possible triangulation of the torus is given by taking $G$ to be the Heawood graph which has $14$ vertices.
However, famously, the dual $G^*$ is $K_7$, the complete graph on $7$ vertices so we cannot $3$-color its vertices.

A natural family of triangulations to consider is as follows.
Consider an infinite square grid with vertices labeled by elements of $\mathbb{Z}^2$.
Divide each square into two triangles by adding an edge from $(i,j)$ to $(i+1,j+1)$ for all $i,j$.  
Call the resulting triangulation of the plane $\Lambda$,
and color each vertex by color $i+j \bmod 3$.
Let $u=(a,b)$ and $v=(c,d)$ with $a,b,c,d\in\mathbb{Z}$
and triangulate the torus by $\Lambda$ modulo $u,v$.
As a consistency relation on the coloring, we require $a+b=0 \bmod 3$ and $c+d=0\bmod 3$.
The number of qubits of the resulting code is equal to the number of triangles, 
which is equal to twice the number of vertices, and hence equals $2|ad-bc|$.

\begin{figure}
\centering
\includegraphics[width=1in]{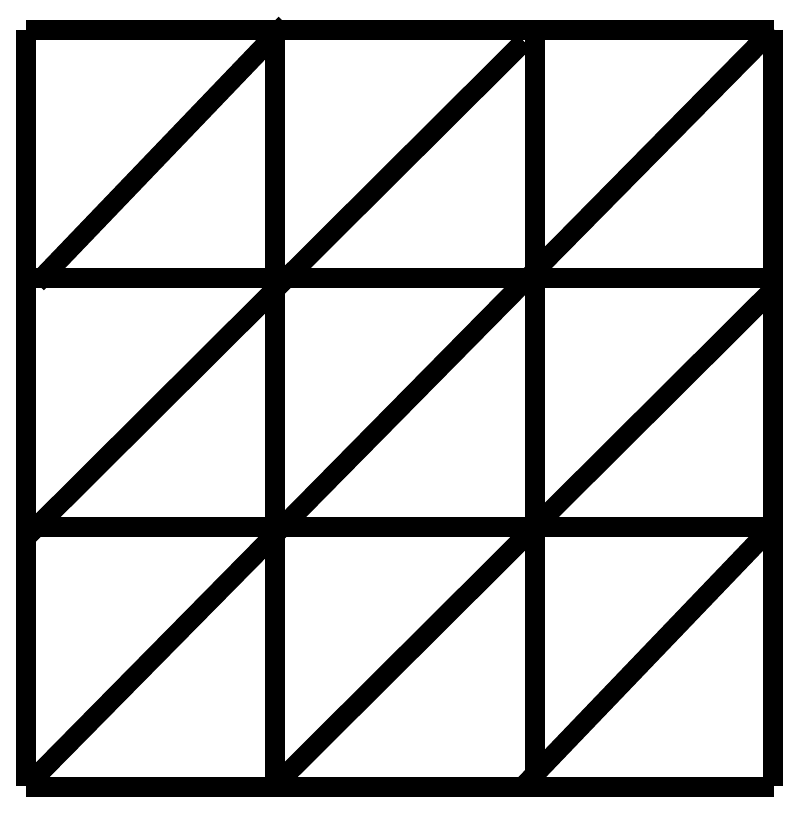}
\caption{
	Dual graph for $18$-qubit code on a torus.
	Left and right sides of figure are identified, as are top and bottom sides.
}
\label{fig:torus18}
\end{figure}

In the particular case where 
\begin{align}
\begin{pmatrix}
a & b \\ c & d 
\end{pmatrix}
=
\begin{pmatrix}
1 & 0 \\ 0 & 1
\end{pmatrix}\ell \quad \text{ where } \ell = 0 \bmod 3,
\label{eq:ad1}
\end{align} 
we have $n=2\ell^2$ qubits.
For $\ell=3$, this gives an $18$-qubit code. 
See \cref{fig:torus18}.

The notion of ``distance'' for codes with dynamically generated logical qubits must be defined carefully.
The reason is that a logical error can occur purely as a result of measurement outcome flips
without any single-qubit Pauli errors occurring.
For simplicity, we just consider the simplest definition of distance:
the minimum number of single-qubit Pauli errors 
which can produce a logical error without causing a nonvanishing error syndrome.
We claim that the distance of this family of codes is $d=(4/3) \ell$, and so these codes have $n=(9/8)d^2$ qubits (interestingly, the same factor of $9/8$ was found for color codes in \cite{bombin2007optimal}, and a $(9/8)d^2$ family of honeycomb codes was noted in \cite{Gidney2021}).
We will examine the decoding graph defined by single-qubit Pauli errors
and count the number of edges in a nontrivial homology cycle.
It suffices to consider one connected component of the decoding graph,
because any single-qubit Pauli error activates at most two edges
that belong to the different connected components.

With the bulk sequence $\ldots,0,1,2,\ldots$ of period~$3$ as in \cite{honeycomb},
the decoding graph is the $1$-skeleton of the simple cubic lattice,
whose (111) direction of the cubic lattice is our time direction.
For a cycle in the $3$d graph, there is a highest point.
Since the $3$d graph does not have any horizontal (orthogonal to the time axis) edge,
the highest point must have two edges that overlap with some square.
We can remove the highest point by changing the error cycle by the boundary of the square, 
reducing the elevation of that point. 
The same can be done to the lowest point in the cycle. 
This modification cannot increase the weight of the chain by construction.
Continue until we arrive at a cycle $C$ along which the time coordinate takes only two values, 
because the boundary of a square assumes three values in time coordinates.
Without loss of generality, suppose these two values of time coordinate correspond to rounds $0,2 \bmod 3$.
Then, the error chain corresponds to a path on the edges of the triangulation (e.g., the edges of \cref{fig:torus18}) 
which avoids type~1 vertices.
This is equivalent to investigating the graph on which the embedded toric code after round~1 lives
and counting the number of edges in nontrivial homology cycles.
Therefore, the code distance is $(4/3)\ell$.

With the modified bulk sequence $\ldots,0,1,2,0,2,1,\ldots$ of period~$6$,
the decoding graph is no longer the simple cubic lattice,
but a similar argument holds.
There are degree-$3$ vertices whose edges are all pointing towards the future or all towards the past.
If an error cycle passes through a degree-$3$ vertex,
then we can make the cycle avoid this vertex by changing the error cycle with a square.
Then, the resulting cycle does not visit any edge connecting a degree-$3$ vertex and a degree-$9$ vertex, and hence lives on a graph consisting of degree-$6$ vertices only.
The argument above applies again.
In conclusion, the code family defined by \cref{eq:ad1} has code distance $(4/3)\ell$
for both of the two versions of the bulk measurement sequence.

Related to our parallelogram is the following code family on torus.
\begin{align}
\begin{pmatrix}
a & b \\ c & d 
\end{pmatrix}
=
\begin{pmatrix}
1 & 2 \\ -1 & 1
\end{pmatrix} m \quad \text{ where } m = 1,2,3,\ldots.
\label{eq:abcd12m11}
\end{align}
The number of qubits is $n = 6 m^2$.
The same reduction of cycles as above holds for this family,
and the code distance (with respect to single-qubit Pauli errors) is $2m$.
This family can realize a finer sequence of code distances than \cref{eq:ad1}.

\bibliographystyle{apsrev4-1}
\nocite{apsrev41Control}
\bibliography{refs}
\end{document}